# Organized Condensation of Worm-Like Chains

H. Schiessel[1,2], J. Rudnick[1], R. Bruinsma[1], and W. M. Gelbart[2]

[1]Departments of Physics and [2]Chemistry and Biochemistry, University of California, Los Angeles, Los Angeles, CA, 90095-1547

## Abstract

We present results relevant to the equilibrium organization of DNA strands of arbitrary length interacting with a spherical organizing center, suggestive of DNA-histone complexation in nucleosomes. We obtain a rich phase diagram in which a wrapping state is transformed into a complex multi-leafed structure as the adhesion energy is reduced. The statistical mechanics of the "melting" of a rosette can be mapped onto an exactly soluble one-dimensional many-body problem.

Reversible coordinated condensation of long eukaryotic DNA strands into a highly compacted package (chromatin), and the controlled swelling of chromatin, are essential requirements for the successful duplication and replication of DNA [1]; strands of the order of a meter are orderly packed into micron-size nuclei without getting tangled-up. It is known that certain enzymes – the topo-isomerases – assist the condensation process by untangling knotted DNA, by releasing the inevitable torsional stresses that are produced during condensation, and by facilitating the "super-coiling" of DNA. It is also known that condensed DNA strands are wrapped around organizing proteins (globular complexes of histones) that carry a charge opposite to that of DNA. Gene transcription is believed to involve some kind of loosening of this "wrapping" state [1].

DNA condensation has recently attracted considerable experimental [2] and theoretical [3] interest from physicists, stimulated in particular by the development of new techniques of single-molecule micro-mechanics [4]. These new tools open up the possibility of detailed mechanical probing of DNA condensation. The study of DNA condensation simplifies significantly if we focus on the interaction of DNA with just one or a few "organizing centers", i.e., particles which *either* condense the DNA strand into a wrapped (continuously adsorbed) state *or* induce it into a centro-symmetric open structure. Even for this simple case we still encounter puzzling results. The *in-vitro* phase diagram of mixtures of short DNA strands with single octameric-histone (nucleosomal [5]) organizing-centers has been studied by Yager et al. [6]. They found an athermal first-order phase transition as a function of the DNA-histone interaction strength (controlled by changing the salt concentration) from a wrapped state to a dissociated state, consistent with the simple ("all or none") unwrapping transition proposed by Marky and Manning [7]. On the other hand, in a numerical study by Wallin and Linse [8] of the association of a long charged polymer ("polyelectrolyte") with an oppositely charged sphere, a more gradual change was observed as the chain stiffness was increased, with one or more loops extending out of the sphere. Finally, recent work by Polach and Widom [9] found that the nucleosome wrapping state actually represents a *dynamical equilibrium* in which wrapped portions of the DNA strand also spent part of their time in a dissociated state. They measured the binding energy of DNA to the nucleosome to be about $0.15$–$0.2 k_B T$ per base-pair under standard conditions.

In this paper we report on a model study of the finite-temperature conformation of DNA (or other semi-flexible polymers) interacting with a spherical organizing center, in an effort to further elucidate the nature of the unwrapping transition. We make use of the popular "Worm-Like Chain" (WLC) model [10] for DNA, which describes the molecule as a semi-flexible tube characterized by two elastic moduli, the bending and torsional stiffnesses. This model is able to describe, on a quantitative basis, the force-extension curve of DNA as measured by micro-mechanical methods [11,12] and it has been used to examine size-reduction of DNA loops by supercoiling, and the effect of thermal fluctuations on supercoiling [13], as well as the complexation of DNA with model nucleosomes [14]. In the present work we map the semi-flexible strand/spherical organizing center system onto an exactly soluble one-dimensional statistical mechanics problem, and exploit the solutions to formulate a phase diagram exhibiting a wide range of wrapped and open structures characterizing the chain-ball complexes.

The elastic energy of a closed WLC loop of length $L$ attached to an adhesive sphere at $N$ sites can be expressed as:

$$H = \tfrac{1}{2}\kappa \oint ds \left[ \left(\frac{1}{R(s)}\right)^2 + C\left(\frac{d\phi}{ds}\right)^2 \right] - \mu N \qquad (1)$$

In Eq. (1), $\kappa$ is the bending stiffness and $1/R(s)$ the curvature of the chain at the point $s$ along its contour. The stiffness is normally expressed as $\kappa = \ell k_B T$, with $\ell$ the orientational persistence length of the chain (about 500Å for DNA under standard conditions). The torsional angle of the chain is $\phi$ and the torsional stiffness is $C$. (For DNA, $C$ is comparable in magnitude to $\kappa$; we will set $C = \kappa$, and then argue later – see below Eq. (3) – that this term can be neglected altogether.) The diameter $b$ of the chain (20Å for DNA) is assumed small compared to the diameter $D$ of the spherical organizing center (about 110Å for nucleosomes). The last term of Eq. (1) describes the interaction between the sphere and the WLC, with $\mu\sqrt{D}$ the binding energy per adhesive site; is the binding energy per unit length, the length of adsorbed chain per site is of order $\sqrt{D}$, and is the range of the attractive interaction. We will assume short-range interactions with $<<b<<D$. We find $\mu \approx k_B T$ if we take for the physiological Debye screening length (10Å), set $D = 110$ Å, and estimate from the measured $0.1$-$0.2 k_B T$ per base-pair [9] (with 3.4Å per base-pair).

The equilibrium number $N^*$ of adhesive sites, and the preferred configurations of the chain, must be obtained by minimization of the free energy corresponding to Eq. (1). We will first search for minima of the elastic energy, neglecting thermal fluctuations.

*A) The Rosette State.* This search can be performed systematically by applying the Kirchhoff theory for thin rods [15]. Kirchhoff theory relates stationary points of the WLC energy (i.e. $H + \mu N$) to a well-studied classical mechanics problem: the trajectory of a spinning top. Different initial conditions on the Eulerian angles of the top correspond to different stationary points of the WLC energies. The solutions can be characterized in terms of the (topologically conserved) "linking number" ($Lk$) of the closed strand. If a loop is constructed by closing a strand on itself, then $Lk$ equals the number of turns imposed on the strand before closing it. Loops of non-zero linking number respond by a combination of twist and spatial distortion – known as "writhing" (DNA supercoiling is a form of writhe).

We impose the following physical conditions on solutions of the Euler-Lagrange equation generated in this manner: (1) The WLC must close on itself. (2) It must be possible to inscribe a sphere of diameter $D$ inside the WLC that touches the WLC at $N$ points. (3) There ought to be no self-intersection of the WLC chain with itself if it is surrounded by a tube of radius $b$. (4) The solution must be stable against small perturbations.

Remarkably, we were able to obtain a family of knotted rosette-shaped WLC stationary points [16] that obey all of these conditions. The rosette can be characterized by the number of loops $N$ of the rosette. For $N = 3$, the rosette has the topology of the trefoil knot. Fig. 1 shows the $N = 5$ rosette (computed numerically). It is self-evident that a sphere (or cylinder) can be inscribed in the central hole of the rosette. For each $N$, we adjusted the linking number of the loop to minimize the elastic energy. For large $N$, rosettes with the optimal linking number have little twist but considerable writhe. By varying the degree of writhe of the solution, a family of solutions was obtained, for given $N$, with different hole diameters $D(N)$. Solutions with small writhe resemble a braided torus and have a large central hole, while solutions with the maximum amount of writhe have the smallest central hole diameter as well as the lowest elastic energy. The inset of Fig. 1 shows the elastic energy of a loop of length $L$, in dimensionless units, for the $N = 5$ rosette state as a function of the degree of writhe. No solutions were found [17] for hole diameters $D(N)$ below a threshold $D_{min}(N) = gL/N^2$ with $g \approx 0.47$. Examination of computed rosette shapes with minimum hole diameters, such as the one shown in Fig. 1, showed that self-intersection at the center of odd $N$-leafed rosettes with a minimum hole diameter is avoided for chain diameters $b$ less than approximately $D_{min}(N)/N$.

It will be demonstrated elsewhere [18] that these solutions actually are *saddle-points* of the WLC energy. The key role of the organizing center is here to stabilize the saddle-point solutions and turn them into true minima of the energy. The energy $E_{min}$ of a minimum-hole rosette depends on $L$ and $N$ as $E_{min}(N) \approx 2A\ N^2/L - \mu N$ with $A=7.02$ (see inset of Fig. 1). A physical interpretation of this result can be based on earlier work by Yamakawa and Stockmayer [19] (YS) who showed that a loop of length $l$, formed by imposing common endpoints on a WLC strand, assumes the form of a lemniscate-shaped leaf with an 81 degrees apex angle. The elastic bending energy of the leaf is $e(l) = 2A/l$ and $E_{min}(N)$ given above just equals $Ne(l/N)$ plus the adhesion energy. We have verified numerically that the leafs of the rosette indeed have apex angles close to 81 degrees. The energy $E_{min}(N)$ exhibits a minimum as a function of the number $N$ of rosette leafs for $N = N^*$ with $N^* \approx (\mu/k_B T)(L/4A)$. We thereby obtain a compact, condensed structure controlled by the adhesion energy and the persistence length that competes with the wrapped state. There are two restrictions.

The optimal number of leafs of the rosette grows as we increase $L$ but the sphere can not accommodate an unlimited number of contact sites without self-intersection (see Fig. 1). From packing considerations, it can be demonstrated that the maximum leaf number $N_{max}$ is of order $(D/b)^{3/2}$.

For larger adhesion energies, we enter the *wrapping state* with the WLC covering the sphere [20]. It is an elementary exercise to show that this happens when reaches the critical value $_c = 2/D^2$. Interestingly, from the earlier numerical estimates for DNA-histone complexation, $2/D^2$ is quite close to $_c$, suggesting that nucleosome-DNA complexes under standard conditions are indeed close to the wrapping transition point.

*B) Melting the Rosette.* The symmetric, organized rosette is only the minimum energy state. This delicate structure might be expected to become unstable in the presence of thermal fluctuations. Thermal fluctuations in fact are well known [10] to disorder minimal energy states of the WLC. To investigate the stability of the rosette against thermal fluctuations, we start from a single, large loop of length $L$ and construct the rosette step by step, by attaching to the organizing center lemniscate-shaped leafs of variable length of the kind examined by YS. The finite-temperature free energy cost $F(l,\ )$ of introducing a single leaf of length $l$ and apex angle into a large strand was computed by YS. Using path-integral methods they found (apart from constants):

$$F(l,\ )/k_B T = \begin{cases} 2A\dfrac{}{l} + \ln l/ + W(\ ) + \ldots & l << \\ \dfrac{3}{2}\ln l/ + \ldots & l >> \end{cases} \qquad (2)$$

The function $W(\ )$ has a minimum when the apex angle of the leaf is approximately 81 degrees. The logarithmic entropic contribution to $F(l,\ )$ imposes a free energy penalty for *large* leafs while the enthalpic $1/l$ contribution imposes an energy penalty for *small* leafs. For given , $F(l,\ )$ has a shallow minimum as a function of $l$ near the persistence length . The total free energy cost $F_N(\{l_i\})$ of introducing an $N$-leafed rosette *for a fixed distribution* $\{l_i\}$ of leaf lengths into a large loop is, then:

$$F_N(\{l_i\}) = \sum_{i=1}^{N} f(l_i) - \mu N$$

$$f(l)/k_B T = 2A\frac{\xi}{l} + \frac{3}{2}\ln\frac{l}{\xi} \qquad (3)$$

In Eq. (3), $f(l)$ constitutes an interpolation formula between the large and small $l$ limits of $F(l,81°)$ as given by Eq. (2) (note that Eq. (3) neglects the twist energy of Eq. (1), which we found to be small for large leaf sizes). The full *free* energy is obtained by performing a Gibbs average over all possible leaf distributions:

$$G(P) = -k_B T \ln \int_0^\infty \prod_{i=1}^{N} \frac{dl_i}{b} e^{-\frac{1}{k_B T}\left[\sum_{i=1}^{N} f(l_i) + P \sum_{i=1}^{N} l_i\right]} - \mu N \qquad (4)$$

with the chain-diameter $b$ acting as a short-distance cut-off. To obtain Eq. (4), we performed a Legendre transform $G = F + PL$ in order to satisfy (on average) the constraint $\sum_{i=1}^{N} l_i = L$; the quantity $P$ is defined by the condition $L = dG/dP$. Physically, $P$ is the overall *tension* of the loops induced by their adhesion to the sphere.

It is interesting to note that $G(P)$ is mathematically identical to the free energy in the constant pressure ensemble of a one-dimensional many-body system of $N$ particles under a "pressure" $P$ confined to a circular track of length $L$. The $N$ particles are interacting via a "nearest-neighbor pair-potential" $f(l)$ while $\mu$ is the "chemical potential" of the particles. According to Eq. (3), the effective pair potential $f(l)$ is concave (i.e., $d^2 f/dl^2 < 0$) for inter-particle spacings exceeding a spinodal threshold spacing of order the persistence length $\xi$. Experience with many-body systems suggests that if the mean spacing between particles exceeds a spinodal threshold, then we should expect *phase-decomposition* [21]. This suspicion is confirmed if we apply mean-field theory (MFT) to evaluate $G(P)$ by assuming a narrow-peaked leaf size distribution. MFT predicts that, in the coexistence regime, the rosette has leaf sizes comparable to the persistence length $\xi$. Because of the physical meaning of the persistence length (the correlation length for the decay of orientational order along the chain), it is certainly reasonable that the physical structure of the rosette changes when the mean leaf size tries to exceed $\xi$.

The free energy Eq. (4) can however be evaluated exactly

$$G(P) = Nk_BT \left[ 2\sqrt{\frac{2A\,P}{k_BT}} + \ln \frac{b/\ell}{\sqrt{\ell/2A}} \right] - \mu N \qquad (5)$$

with a corresponding non-linear "tension-extension" curve $P(L) = 2A\,k_BT(N/L)^2$. (We may also interpret this relation as the non-ideal "equation of state" of the equivalent many-body system.) Using Eq. (5) and $P(L)$ to compute the free energy $F = G - PL$, we recover the "$T=0$" result $E_N(\min)$ discussed above provided we replace $\mu$ by $\mu^* = \mu - k_BT\ln(\sqrt{2A}\,b/\sqrt{\ell})$ and the same holds for the formula for the optimal number of leaves $N^*$.

Using Eqs. (4) and (5), it is straightforward to compute the first and (reduced) second moments of the leaf size distribution: $<l> = L/N$ and $\sqrt{<(l-<l>)^2>}/<l> = \sqrt{L/4AN}$. Thus we encounter *no* phase-coexistence: the leaf size grows with leaf length in the same manner as the "$T=0$" solution. This does *not* mean however that the rosette structure is not altered when the mean leaf size $<l>$ exceeds $\ell$, because the reduced second moment starts to exceed one at that point (actually, at $<l>$ exceeding $4A\ell$). We can therefore identify $L/N \approx \ell$ as the onset point of *heterogeneity* of the leaf size distribution; the orderly, symmetric rosette is starting to "melt". It must be emphasized though that $G(P)$ is analytic and that there is no true thermodynamic singularity.

This heterogeneous, swollen rosette state is very *fragile*. If we redo the calculation *after cutting the closed loop*, i.e., allowing for two free ends, we recover the result of the closed loop case but only for $L/N << A\ell N$. In the opposite limit, $L/N >> A\ell N$, we find $<l> \approx \sqrt{A\ell L/2}$ and $\sqrt{<(l-<l>)^2>}/<l> \approx (L/A\ell)^{1/4}/2$. The mean leaf size is now small compared to $L/N$ and most of the WLC is part of two large chains that emerge from the rosette. This resembles the appearance of phase-coexistence as predicted by MFT, but note that MFT does not correctly predict the *onset point* of the quasi-phase coexistence, giving $L/N \approx \ell$ rather than $\ell N$. It appears surprising that cutting the loop has such a dramatic effect on the rosette structure. Although we have not verified this, we expect that even a very small external tension on a closed rosette would cause it to collapse once $L/N \gg N$.

We have summarized our results in a "phase-diagram" of total loop length $L$ versus the adhesion energy μ (see Fig. 2) for the case $N_{max} = 5$. The full lines indicate transition points between rosettes with a different number $N^*$ of leafs as computed from Eq. (4). The thin dashed horizontal lines indicate the melting points $L/ \sim N^*$. Note that one encounters phase separation for $L/ \gg A(N^*)^2$ in the case of an open loop. The vertical gray bar marks the wrapping transition.

We also investigated the failure of MFT by formally re-computing the free energy for dimensions different from $d = 3$ (for closed loops). For $d < 4$, the mean loop size is $L/N$, while for $d > 4$ we find $<l> \sim 8A / (d-4)$ (for large $L$), consistent with MFT; the effective *compressibility* of the equivalent many body system becomes negative in that case, indicative of phase-separation. This result is suggestive that $d = 4$ acts somewhat as an "upper-critical dimension", with MFT correct above $d = 4$. The interaction between a WLC and a sphere is thus an interesting problem of many-body physics in general.

We now return to the experiments described in our introduction. Suppose the DNA strand lengths $L$ are comparable to the persistence length    and suppose we vary the adhesion energy μ for *fixed*    (and $L$), then we perform a trajectory in the phase-diagram along a horizontal line. For that case Fig. 2 predicts a simple "first-order" transition (upon decreasing μ) from a wrapping state to one where there is just a single contact between the loop and the organizing center, consistent with the findings of Yager et al. [6] in which unwrapping occurs in a single step upon the addition of salt. If, on the other hand, for a given large loop we decrease    for *fixed* (*sub-critical*) μ, starting from a large value of    (i.e., small enough $L/4A$ ), then the trajectory is a vertical straight line upwards in Fig. 2. One encounters a complex sequence of rosette states with different $N^*$, consistent with the simulations of Wallin and Linse [8]. An interesting test of the theory would be to re-do the experiments of Yager et al. [6] for circular DNA plasmids with $L \gg$   . We predict that rosettes should form as the adhesion strength is decreased, as described in the first scenario immediately above. The resulting size of the rosettes might be monitored by electrophoresis and the effect of cutting the rosette could be simulated with topo-isomerases. A more general biological relevance of the fragile rosette state could be as a swollen state in which genes are freely accessible to transcription factors and other proteins while the DNA still holds on to the nucleosomes for future re-compaction (with the residual binding energy $\mu N_{max}$ being sufficient for this process).

We would like to thank Stella Park and Jens-Uwe Sommer for useful discussions. We would like to thank the NSF for support under grant # DMR-9706646.

*Figure Captions:*

Figure 1: Five-leaved rosette with minimal central-hole produced by solutions of Euler-Largrange equations. The spherical organizing center is shown. Inset: elastic energy of the $N = 5$ rosette in dimensionless units ($<l> \quad L/N$) as a function of the degree of writhe. The lowest energy corresponds to the minimal hole size.

Figure 2: Phase diagram as a function of the length L of the loop and the site adhesion energy μ. The wrapping transition is shown as a grey vertical bar. The horizontal dashed lines are points beyond which the rosette leaf structure is heterogeneous due to thermal fluctuations. For large loop lengths, the rosette "phase-separates" if the loop is cut.

Fig.1

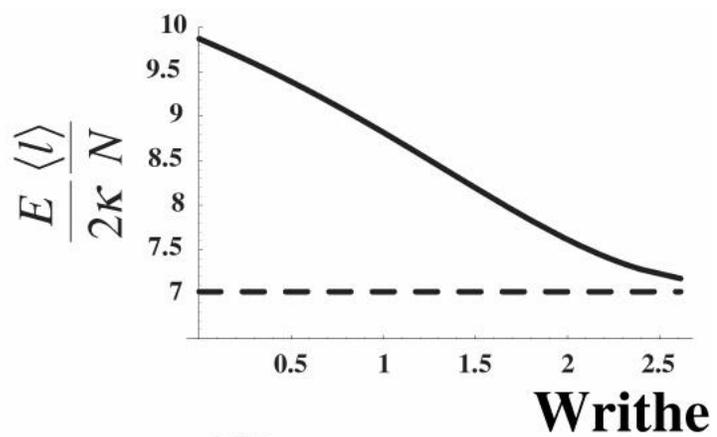

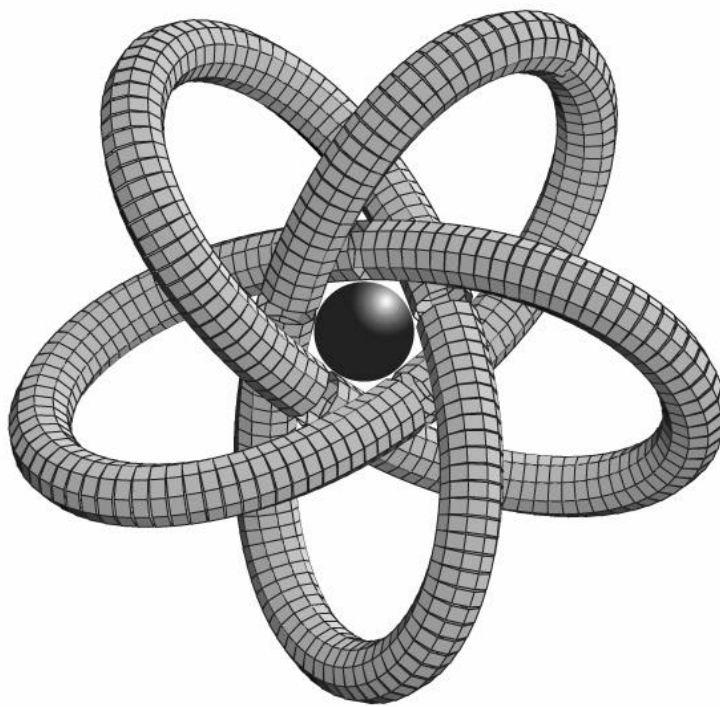

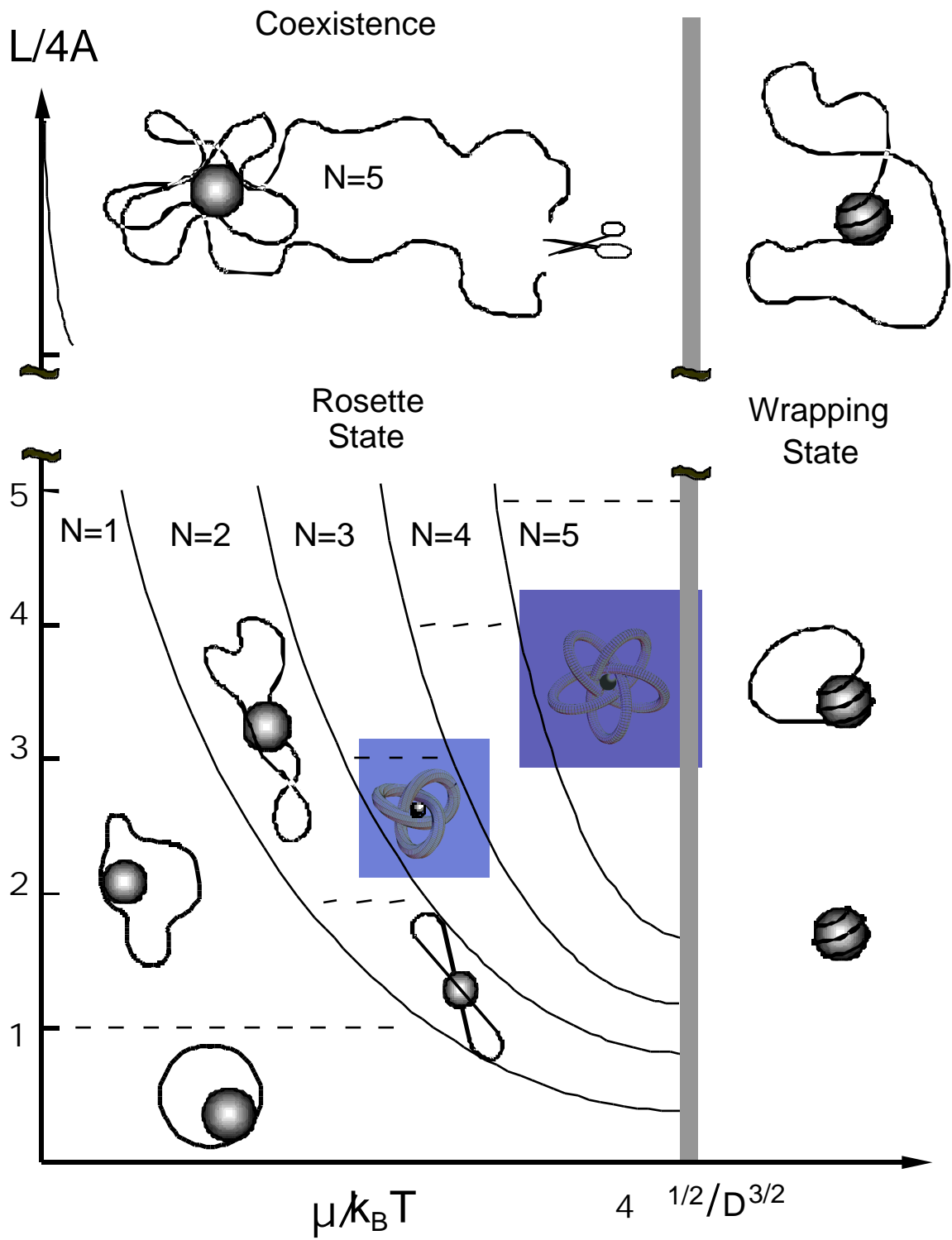

[21] Decomposition of a two-loop polymer with a slip-link was proposed by J.-U. Sommer, J. Chem. Phys. **97**, 5777 (1992).